\documentclass[referee]{raa}            

\usepackage{graphicx,times}             
\usepackage{epstopdf}
\usepackage{natbib}
\usepackage{amssymb,amsmath}
\bibpunct{(}{)}{;}{a}{}{,}

\usepackage[a4paper=true,dvipdfm=true,pagebackref=true]{hyperref}
\hypersetup{colorlinks = true, linkcolor = green, anchorcolor = red, citecolor = blue, filecolor = red, pagecolor = red, urlcolor = red}

\begin{document}

   \title{CAN INJECTION MODEL REPLENISH THE FILAMENTS IN WEAK MAGNETIC ENVIRONMENT?}

   \volnopage{Vol.0 (20xx) No.0, 000--000}      
   \setcounter{page}{1}          

   \author{Peng Zou
      \inst{1,2}
   \and Chaowei Jiang
      \inst{1}
   \and Fengsi Wei
      \inst{1}
   \and Wenda Cao
      \inst{3}
   }

   \institute{Institute of Space Science and Applied Technology, Harbin Institute of Technology,      Shenzhen 518055, China; {\it awaken@hit.edu.cn}\\
        \and
             School of Astronomy \& Space Science, Nanjing University, Nanjing, 210023, China\\
        \and
             Big Bear Solar Observatory, New Jersey Institute of Technology, ~40386 North Shore Lane, Big Bear City, CA 92314, USA\\
\vs\no
   {\small Received~~20xx month day; accepted~~20xx~~month day}}

\abstract{We observed an H$\alpha$ surge occurred in the active region NOAA 12401 on 2015 August 17, and discuss its trigger mechanism, kinematic and thermal properties. It is suggested that this surge is caused by a chromospheric reconnection which ejects cool and dense material with the transverse velocity about 21-28 km s$^{-1}$ and the initial doppler velocity of 12 km s$^{-1}$. This surge is similar to the injection of newly formed filament materials from their footpoints, except that the surge here occurred in a relatively weak magnetic environment of ~100 G. Thus we discuss the possibility of filament material replenishment via the erupting mass in such a weak magnetic field, which is often associated with quiescent filaments. It is found that the local plasma can be heated up to about 1.3 times of original temperature, which results in an acceleration about -0.017 km s$^{-2}$. It can lift the dense material up to 10 Mm and higher with a inclination angle smaller than 50$^{\circ}$, namely typical height of active region filaments. But it can hardly inject the material up to those filaments higher than 25 Mm, namely some quiescent filaments. Thus we think injection model does not work well in the formation of quiescent filaments.
}

   \authorrunning{Zou, P. et al. }            
   \titlerunning{Observation of An H$\alpha$ Surge }  

   \maketitle

%
%
\section{Introduction}           
\label{intro}

Solar filaments are heavy and cold objects in the Sun's hot corona and they are closely related with solar eruptions such as CMEs \citep{chen11}. It is commonly believed that filaments are sustained by the special magnetic configurations, such as the magnetic flux ropes or sheared magnetic arches with dips \citep{ks57,kr74}. Their formation process was discussed by many authors \citep{van89,rust94,lizhang16,xia14,xia16,song17}. Three models for replenishing the filament materials are widely accepted, the magnetic flux rope lifting model \citep{rust94,deng00,okamoto08,leake13}, the evaporation and condensation model \citep{mok90,antiochos91,dahlburg98} and the cool material injection model \citep{chae03,liu05,zou16,zou17}. The injection model, which follows the scenario that cool and dense chromosphere materials inject into the filament channel via a chromospheric magnetic reconnection, was firstly observed and proposed by \citet{chae03}. In the observation of \citet{liu05}, they suggested that erupting H$\alpha$ surges may directly link to the replenishment of filament materials. Furthermore, more and more observational evidences support this scenario \citep{zou16,zou17,wang18}. All of these works indicate that a chromospheric eruption can be a source origin of filament replenishment.

Fan-like H$\alpha$ surges, which are recognised as chromospheric eruptions, was reported by several authors in past decades \citep{roy73,asai01,shimizu09,robustini16,li16,hou16,yang16,robustini17}. Most of them are observed above the light bridges of sunspots \citep{roy73,asai01,shimizu09,hou16,robustini16,yang16,robustini17}. They can last for several hours or even one day. During the period, they can be recurrent and exhibit like a wall \citep{yang16}. The speed of their eruption can reach 100 - 200 km s$^{-1}$ \citep{robustini16}. Accompanied with the eruptions, strong intensity enhancements can be observed in footpoints of surges using chromospheric line. Since it is located in polarity inversion lines with the high current density, these brightenings are thought to be indicators of magnetic reconnection \citep{shimizu09}. And the magnetic reconnection can lift the dense plasma up into the corona by magnetic tension force, which was simulated by \citet{jiang11}.

According to the simulation, few papers discussed the replenishment of filament material via injection model \citep{zou16,zou17}. In the study of \citet{zou17}, they showed the extremely brightenings observed in H$\alpha$ images and the co-spatial brightenings in C II and Si IV observed by IRIS. Both of them indicate the violet heating occurred. Associating with the eruptive velocity of filament fibrils, the material replenishment is possible. In contrast, the study of \citet{zou16} showed that the bright points in H$\alpha$ are gender and the transverse velocity of fibrils are small, namely 5 - 10 km s$^{-1}$. It seems the plasma can hardly replenish the filament. However, the filament was forming. It is noted that the observations of filament formation via injection process are all active region filaments and intermediate filaments \citep{chae03,zou16,zou17,wang18}, which have the low magnetic configuration. Thus an interesting question is: can the injection model also work in the weak magnetic environment to replenish the materials for quiescent filaments?

In this paper, we report the observation by the Fast Imaging Solar Spectrograph \citep[FISS,][]{chae13} of the New Solar Telescope \citep[NST,][]{cao10,goode12} in Big Bear Solar Observatory (BBSO) of a fan-like H$\alpha$ surge caused by weak parasitic magnetic field emergence, in order to shed some light on this question.

\section{Observations}
\label{Obs}

\begin{figure}
\centering
\includegraphics[width=10cm]{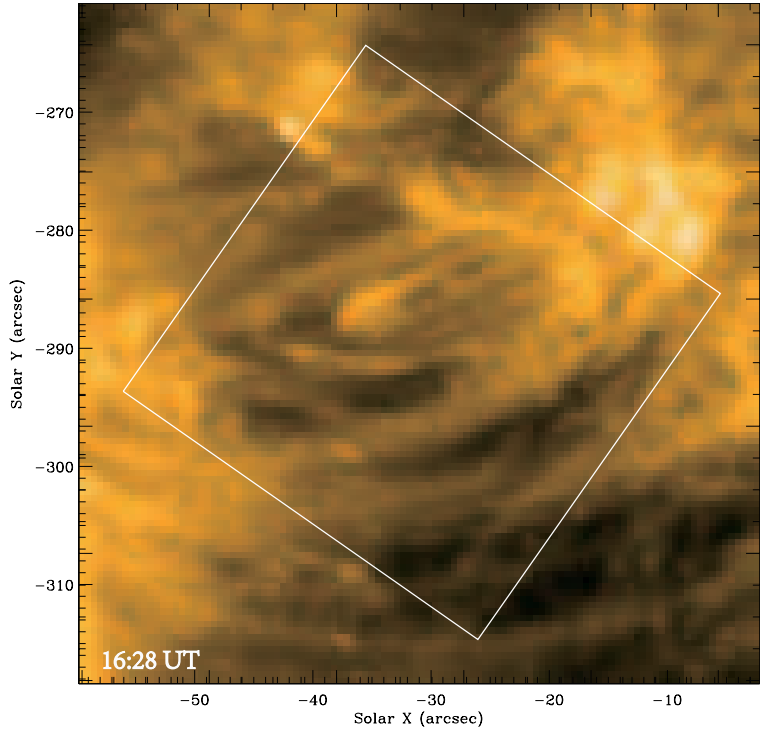}
\caption{Figure shows the 171 \AA\ image at 16:28 UT and the white quadrangle shows the FOV of FISS/NST.}
\label{f1}
\end{figure}

The fan-like surge occurred in active region NOAA 12401 on 2015 August 17. The FOV of FISS is shown in Figure \ref{f1} using white quadrangle. An eruption is well observed by FISS in H$\alpha$ and Ca II spectrum. The FISS is an imaging spectrograph on board the first facility-class solar telescope NST. It adopts an Echelle disperser with field scanning method. Two spectral bands (H$\alpha$ and Ca II 8542 \AA) are available simultaneously in two-dimensional spectra and images. The FISS can provide reconstructed images of photosphere to chromosphere. It can help us understanding the physical process of various phenomenons, such as spicules, prominence and chromospheric eruptions, in multi-layers. The field of view (FOV) of reconstructed images is $40\arcsec \times 40\arcsec$, it is $40\arcsec$ in the slit direction by 250 steps in the scan direction. The spatial resolution is $0 \farcs 16$, the cadence is 43s for both lines and the spectral resolution is 0.0168 \AA\ per pixel for H$\alpha$ and 0.0161 \AA\ per pixel for Ca II 8542 \AA\ respectively. In order to understand the underlying mechanism of this surge, we obtain the magnetograms taken by the Helioseismic and Magnetic Imager \citep[HMI,][]{sch12,schou12} on board the Solar Dynamo Observatory ({\it SDO}). The magnetogram has a spatial resolution of $0 \farcs 5$ and a cadence of 45s. For co-aligning the images taken by different instruments, the reconstructed images of H$\alpha$ line wing are compared with the continuum of HMI/{\it SDO}. For a comparison with the chromospheric images, the 1600 \AA\ images taken by Atmospheric Imaging Assembly \citep[AIA,][]{lemen12} are employed in our work.

\section{Results}
\label{results}

\begin{figure}
\centering
\includegraphics[width=14cm]{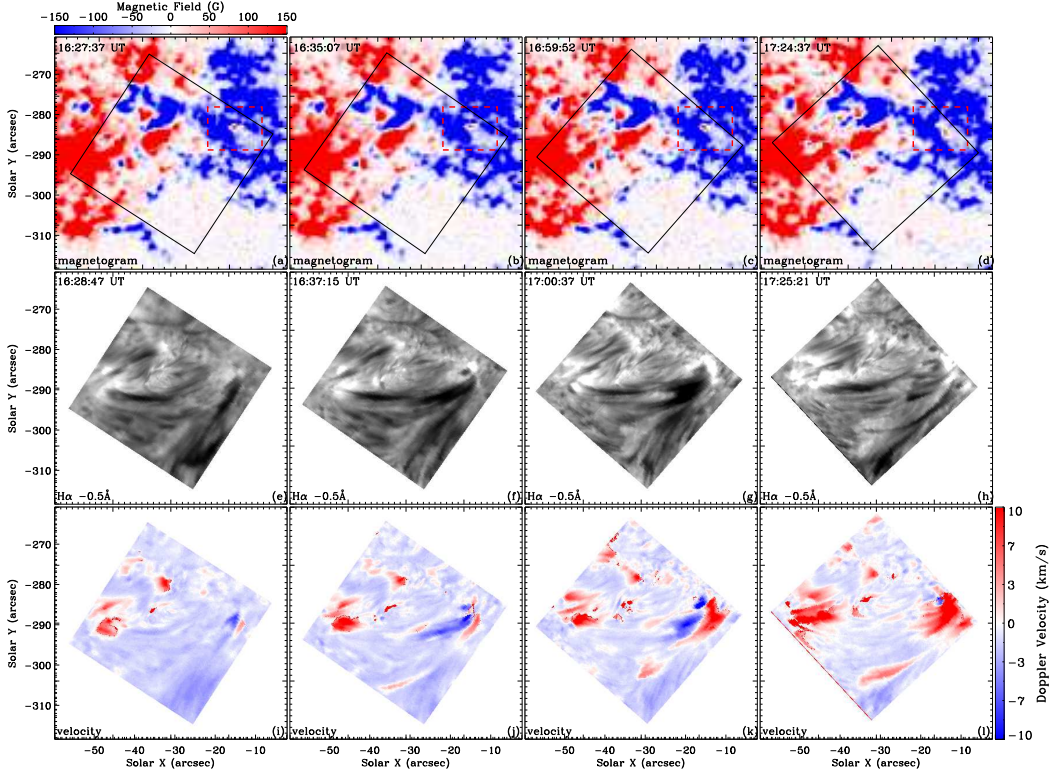}
\caption{Figure exhibits the magnetograms (upper row), H$\alpha$ -0.5 \AA\ images (middle row) and doppler velocity maps (bottom row) respectively. The black quadrangles in magnetograms indicate the FOV of FISS/NST and the red dashed boxes in first row are the area for calculating the positive magnetic flux evolution.}
\label{f2}
\end{figure}

Our observation period is from 16:22 UT to 18:00 UT. However, because of the rotation of the Field of View (FOV), we cannot observe the footpoints of this surge after 17:39 UT. This surge began a little earlier than our observation, thus the initial eruption was existed. After the first eruption, nearby plasmas start brightening and some dark fibrils are elongated from the brightenings. Initially, the brightenings are isolated and intermittent. After 16:45 UT, the brightenings form a ribbon and the dark plasmas recurrent associated the brightenings. We display some newly erupted fibrils at different time in the middle column of Figure \ref{f2} using H$\alpha$ -0.5 \AA. The doppler velocity maps, evaluated by bisector method, indicate that the dark fibrils connected to the brightenings are erupting plasma (see the bottom row of Figure \ref{f2}). The attached H$\alpha$ movie shows the full evolution of the observation period. It is noted that the intensity enhancement of bright points is extremely strong, which is roughly 1.5 or even 2 times of the quiescent area. By contrast, the eruptive fibrils are absorptive in both chromospheric lines and EUV (see in Figure \ref{f1}).

\subsection{Evolution of magnetic fields}
\label{mag}

\begin{figure}
\centering
\includegraphics[width=12cm]{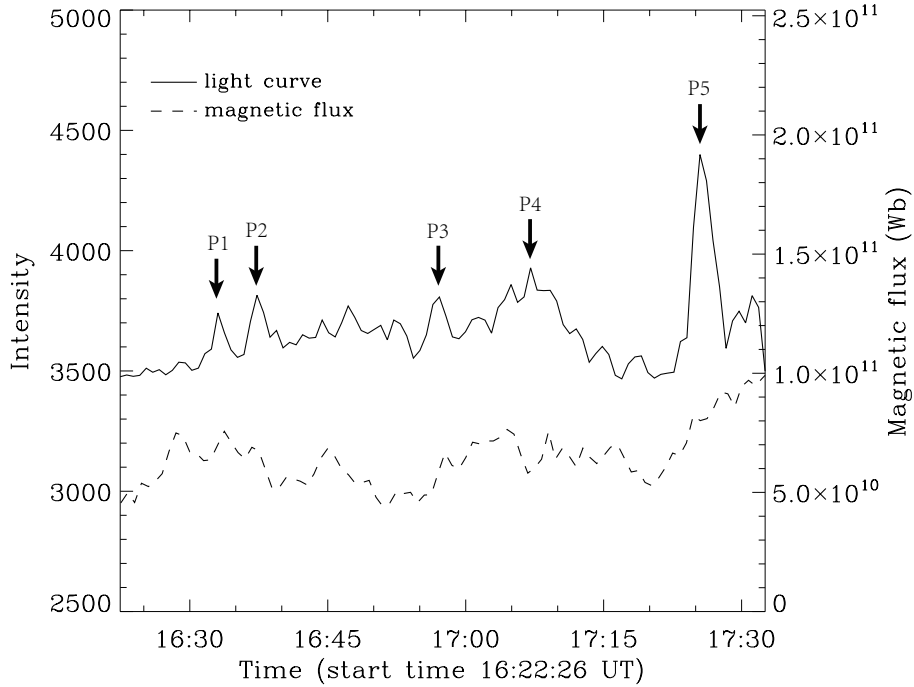}
\caption{The light curve of brightenings observed in H$\alpha$ line center (black solid line) and the positive magnetic flux of the same area (black dashed line).}
\label{f3}
\end{figure}

In order to know whether it is a reconnection-triggered eruption, we also display the magnetograms above the H$\alpha$ image panels in Figure \ref{f2}. It can be seen that the footpoints of this surge are close to the positive parasitic magnetic poles embedded in the negative field. However, some parasitic magnetic poles are so weak (under 100 G) that hardly detected in figure. For illustrating the relationship between the magnetogram and the brightenings, we calculate the light curve of the brightenings and positive magnetic flux of the magnetogram (calculated from the red dashed boxes in Figure \ref{f2}). As we shown in Figure \ref{f3}, the magnetic flux curve exhibits a similar trend with light curve. It is noted that, there are some obvious peaks, which can find a similar peak in magnetic flux, can be seen in the light curve, i. e., the double peaks from 16:31 UT to 16:40 UT in light curve and the similar double peaks from 16:27 UT to 16:37 UT in magnetic flux. Moreover, the fluctuation from 16:58 UT to 17:10 UT in light curve has a similar fluctuation in magnetic flux from 16:54 UT to 17:06 UT as well. In addition, a common point can be seen in both these two time period, i. e., the magnetic flux increases are both four minutes before the increases of the light curve. The relations between the light curve and the magnetic flux imply that the intensity enhancement in Ha is probably caused by the increasing photospheric magnetic flux. All these phenomenon, the brightenings observed by chromospheric lines co-spatial with polarity inversion line (PIL) and the increasing intensity associated with increasing magnetic flux, always indicate the local chromospheric reconnection \citep{isobe07,nelson13,hong17}.

\subsection{Velocities of Surge}
\label{vos}

\begin{figure}
\centering
\includegraphics[width=14cm]{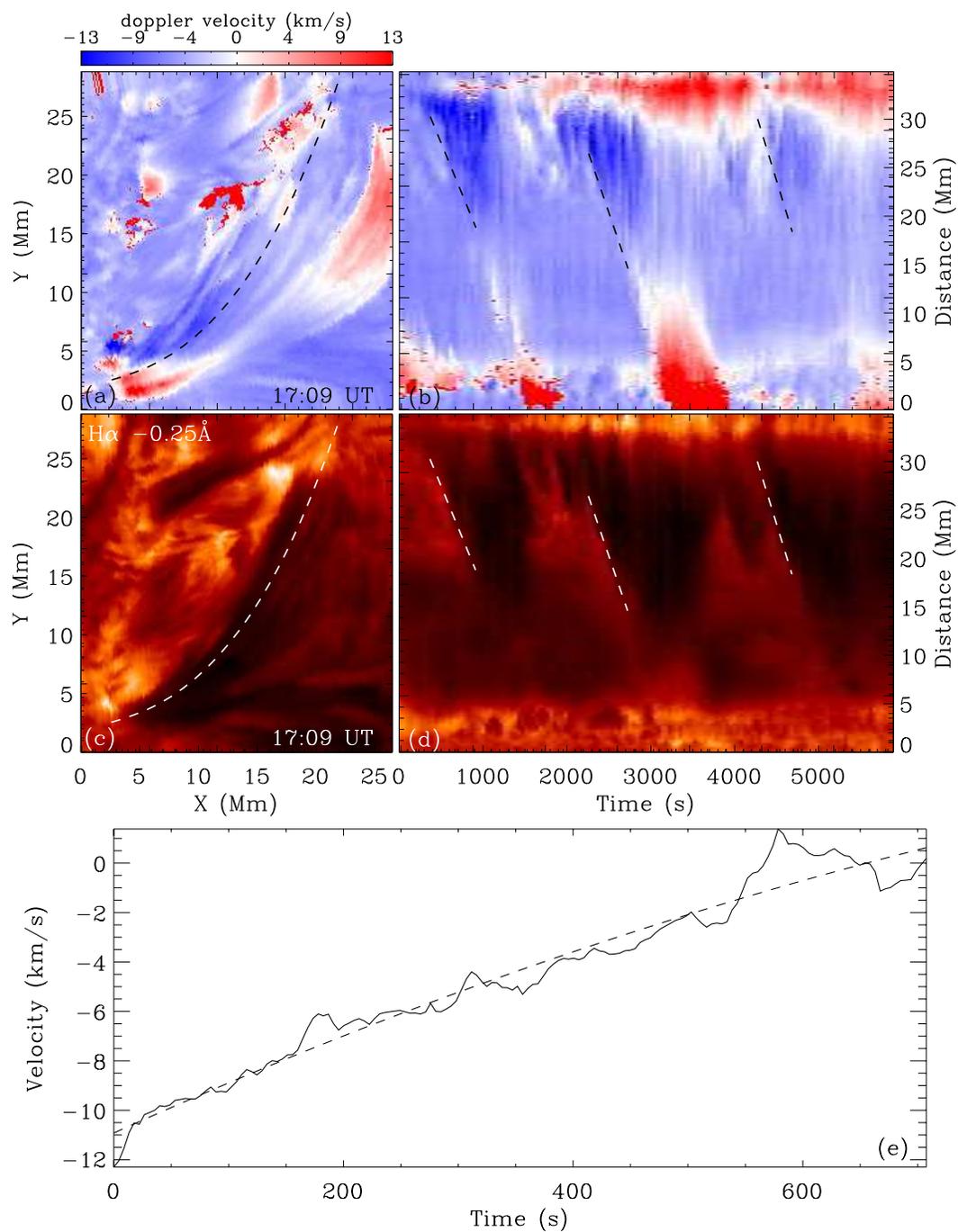}
\caption{Panel (a) and (c) show the selected fibril for calculating the time-distance map in doppler velocity map and H$\alpha$ image respectively. Panel (b) displays time evolution doppler velocity of this slice and Panel (d) is the time-distance map. Panel (e) shows the doppler velocity along one of the erupting fibril.}
\label{f4}
\end{figure}

As we mentioned in section \ref{results}, the erupted plasmas exhibit absorptive in both EUVs or typical chromospheric lines, thus it was recognised as a chromospheric activity. The chromospheric ejections often have a lower eruptive velocity. We evaluate the velocity of this surge including both light-of-sight (LOS) velocity and transverse velocity, via spectrum analysis and time-distance method. The LOS velocity, as shown in the bottom row of Figure \ref{f2}, is about 10 - 13 km s$^{-1}$ for newly formed fibrils and has a rapidly decrease trend along the erupted fibrils. After 16:42 UT, the drop back plasma dominate the area near footpoints. For evaluating the transverse velocity, a time-distance slice of a typical ejective fibril is shown in Figure \ref{f4} panel d, which shows that the speed is about 21 - 28 km s$^{-1}$. We further evaluate the acceleration in LOS direction. In order to evaluate the acceleration, we select one fibril (the same fibril with the one selected for calculating transverse velocity) and calculate its time evolution doppler velocity map (Figure \ref{f4} b). Obviously, some eruption processes can be seen in both time evolution doppler velocity map and time-distance map. Basing on the time evolution doppler velocity map, the doppler velocity evolution of eruption frontiers are selected frame by frame and exhibited in Figure \ref{f4} panel e. The evolution curve gives an acceleration of -0.017 km s$^{-2}$. This acceleration is comparable to the value observed in filament download motion \citep{chae08}.

\subsection{Cloud Model}\label{ecm}

The chromospheric spectrum inversion can provide the information of local or whole chromosphere. In order to inverse the spectrum, \citet{beckers64} proposed a simple method named Becker's cloud model (BCM). However, their method is useful for evaluating the properties of objects above the chromosphere, such as filaments. However, in our observation, the local heating is embedded in the chromosphere. For suiting the embedded objects, \citet{chae14} improved their method, namely embedding cloud method (ECM), and achieve some useful information from their FISS data. Furthermore, using this method, \citet{hong14} also calculated the temperature of the Ellemen bomb from the H$\alpha$ line observed by FISS, which is similar to the results of \citet{fang06} via semi-empirical model.

In BCM, the contrast profile can be described as follows:
\begin{equation}\label{e1}
C_{\lambda}=\left(\frac{S}{I_{\lambda,in}}-1\right)\left[1-exp\left(-\tau_{\lambda}\right)\right].
\end{equation}
Where $C_{\lambda}$ is contrast profile, S is the source function of calculated cloud, $I_{\lambda,in}$ is incidence ray determined by observation of quiescent area $I_{\lambda,in}=R_{\lambda,obs}$ and $\tau_{\lambda}$ is the optical thick of cloud derived from follow equation:
\begin{equation}\label{e2}
\tau_{\lambda}=\tau_{0}exp\left[-\left(\frac{\lambda-\lambda_{1}}{W}\right)^{2}\right].
\end{equation}
Where $\tau_{0}$ is the optical thick of line center, $\lambda_{1}$ is the doppler velocity of cloud and W is the doppler width. According to equation \ref{e1}, the equation has four free parameters: S, $\tau_{0}$, $\lambda_{1}$ and W. When the cloud is considered as an embedded one, then the contrast profile can be derived from follow equation:
\begin{equation}\label{e3}
C_{\lambda}=\left(\frac{s}{R_{\lambda,obs}}-1\right)\left[1-exp\left(t_{\lambda}-\tau_{\lambda}\right)\right] +\left[1-exp\left(-\tau_{\lambda}\right)\right]\frac{S-s}{R_{\lambda,obs}}.
\end{equation}
Here comes some new parameters, where s is the ensemble-average of S and $t_{\lambda}$ is the optical thick of reference area derived by the equation as follow:
\begin{equation}\label{e4}
t_{\lambda}=t_{0}exp\left[-\left(\frac{\lambda-\lambda_{2}}{w}\right)^{2}\right],
\end{equation}
with the three new parameters: $t_{0}$, $\lambda_{2}$ and w.

\begin{figure}
\centering
\includegraphics[width=10cm]{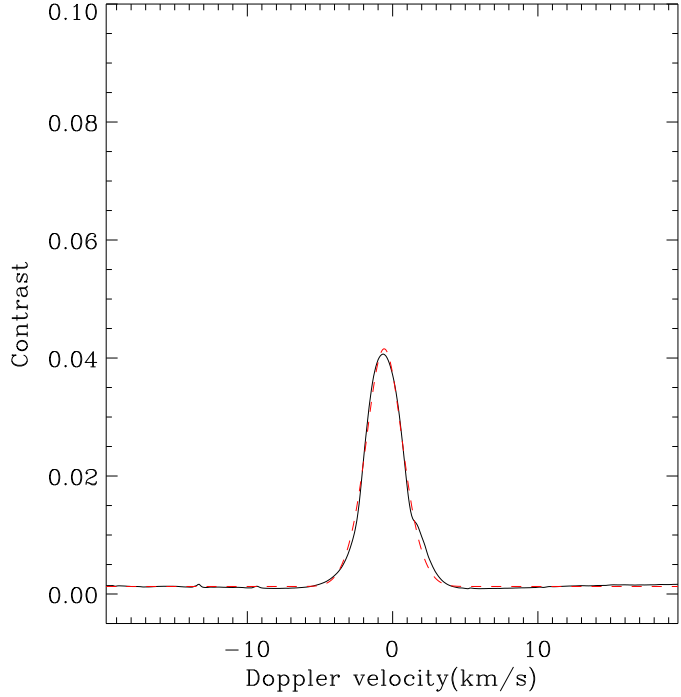}
\caption{Observed rms contrast profile (black solid line) and the fitting curve (red dashed line).}
\label{f5}
\end{figure}

\begin{figure}
\centering
\includegraphics[width=14cm]{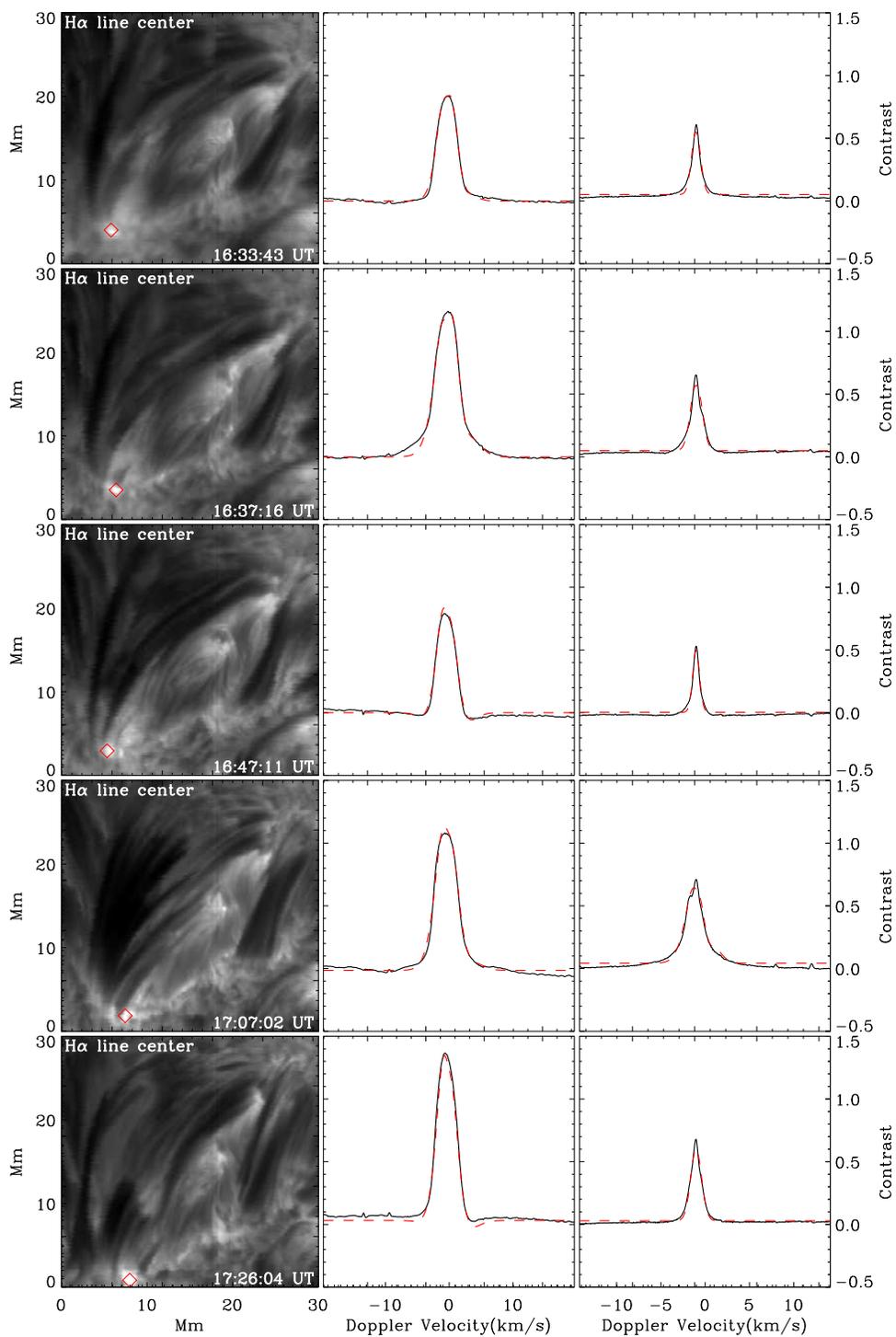}
\caption{First column shows the position of selected brightenings (red quadrangles) respectively. Second column shows the H$\alpha$ line profiles (black solid lines) and fitting curves (red dashed lines) respectively. Third column shows the Ca II line profiles and fitting curves as same as H$\alpha$ line.}
\label{f6}
\end{figure}

\begin{figure}
\centering
\includegraphics[width=10cm]{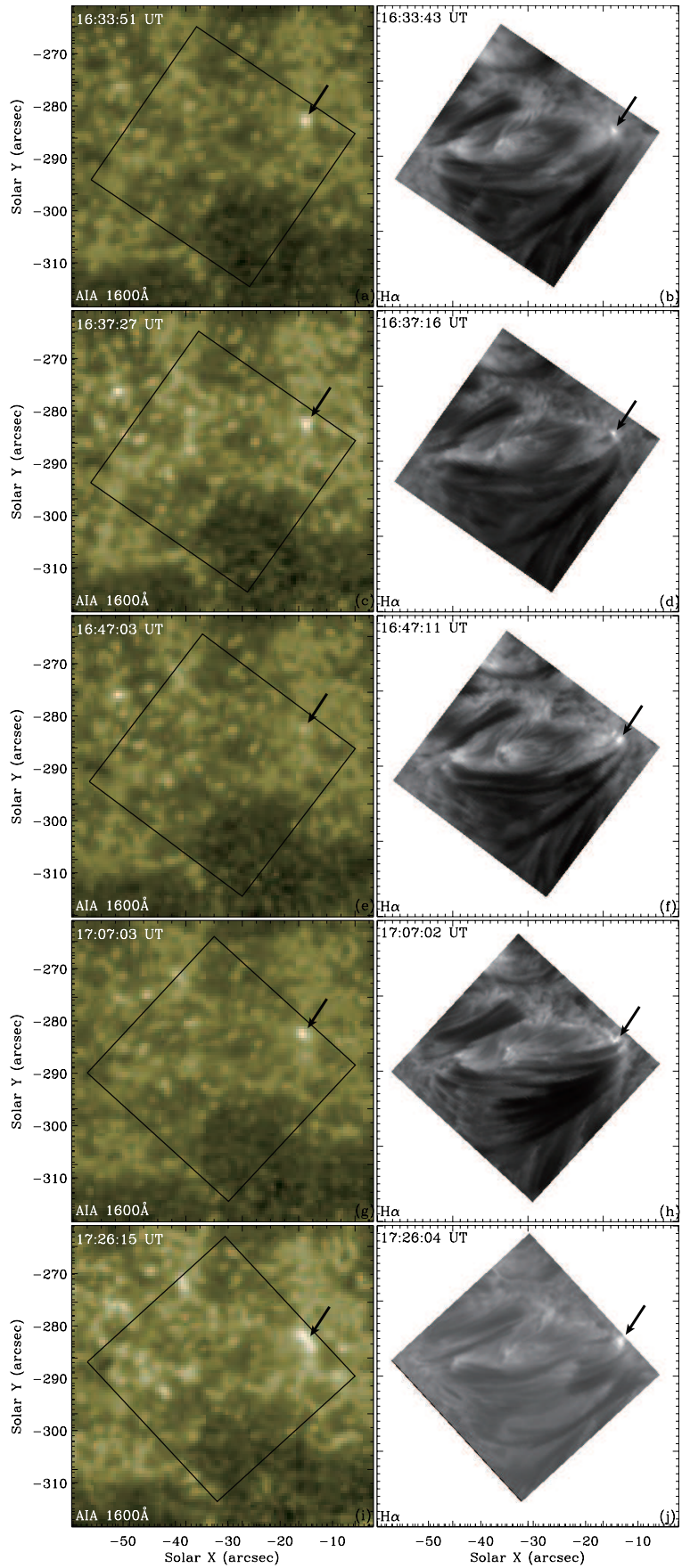}
\caption{Comparison of the 1600 \AA\ images and H$\alpha$ line center images. The left column is 1600 \AA\ images and the right column is H$\alpha$. The black quadrangles show the FOV of FISS/NST in different time respectively. Black arrows indicate the bright point in 1600 \AA\ and H$\alpha$ images}
\label{f7}
\end{figure}

According to the equation \ref{e3}, eight parameters are needed for fitting contrast profile of target objects. To reduce the free parameters, the method mentioned in Appendix of \citet{chae14} is used for fixing four parameters: s,$\lambda_2$, $\omega$ and $t_{0}$. One of the rms contrast profile $G_{\lambda}$ and its fitting curve are displayed in Figure \ref{f5}. According to the assumption of complete frequency redistribution, the source function for H$\alpha$ is calculated as follow:
\begin{equation}\label{e5}
S_{\lambda}=\frac{2hc^2}{\lambda^5}\frac{1}{\frac{b_2}{b_3}exp\left(\frac{hc}{\lambda kt}\right)-1}.
\end{equation}
In this equation, the $\lambda$ is the wavelength of H$\alpha$, the h, k and c are Planck constant, Boltzmann constant and speed of light respectively, the $b_2$ and $b_3$ are departure coefficients of hydrogen atom at energy levels 2 and 3 respectively and T is the local temperature. Since the contrast profiles exhibit a single peak, the main heating is concentrated in H$\alpha$ line center. Thus we choose the heated layer is the height range from 1900 km to 2100 km. And the values of $b_2$, $b_3$ and T will choose from VAL C model. For evaluating the source function of heated footpoints, we select five typical brightenings and signed them in the Figure \ref{f3}. All of them are the peaks of intensity enhancement show in Figure \ref{f3}. The positions (first column), H$\alpha$ contrast profiles and fitting curves (second column) and Ca II profiles and fitting curves (third column) of bright points are shown in Figure \ref{f6}. Basing on the fitting parameters, the relative increase of source function and increased temperature are shown in Table \ref{t1}. The increase of temperatures are about 800K to 3500K, which means a heating up to 8kK - 13kK. This result is a bit lower than the inversion of \citet{robustini17}, namely 14kK. Furthermore, in their study, the intensity enhancement is co-spatial with the 1600 \AA\ brightening. Thus they suggest that the local heating may heat the footpoint up to a higher temperature and the Ca II based inversion is insensitive to temperature above 15 kK because the Ca II is ionized. In comparison, we display the comparison of the H$\alpha$ images and 1600 \AA\ images in Figure \ref{f7}. As we shown in figures, the brightenings in 1600 \AA\ are not co-spatial with the H$\alpha$ one until 17:26 UT. Few minutes after this time, the footpoints of the surge exceed our FOV, thus the further situations are unknown.

\begin{table}
\centering
\caption{The increases of source function and temperature derived from fitting.}
\begin{tabular}{ccc}
\hline
\hline
Point No.&Increased Rate&Increased Temperature (K)\\ \hline
P1&0.44 - 0.82&850 - 2775\\
P2&0.52 - 0.90&950 - 3000\\
P3&0.38 - 0.76&760 - 2560\\
P4&0.40 - 0.77&780 - 2610\\
P5&0.62 - 1.05&1170 - 3480\\ \hline
\label{t1}
\end{tabular}
\end{table}

\section{Discussion and Conclusion}
\label{dis}

Basing on the analysis of a fan-shaped surge in the active region NOAA 12401 and we obtain its properties: The surge has an transverse velocity about 23 - 28 km s$^{-1}$ and an initial LOS velocity about 11 - 13 km s$^{-1}$. The eruptive fibrils are associated with the footpoint brightenings after the increase of magnetic flux. Thus the most probably trigger mechanism of the surge is magnetic reconnection. The surge is absorptive in both chromospheric lines and EUVs. Thus the erupted plasma should be cold and dense, which means this reconnection occurs in chromosphere.

Plasma heating via chromospheric reconnection has been discussed for decades, but it is still unclear whether the plasma can be heated up to tens thousand kelvin. In previous works, some authors suggested that the chromospheric brightenings caused by reconnection can co-spatial with some UVs and EUVs brightenings, such as iris bombs \citep{vissers13,tian16}. However, as we mentioned in section \ref{ecm}, no EUVs or UVs brightenings co-spatial with our observation from 16:22 UT to 17:26 UT (see the black arrows in Figure \ref{f7}). And, during this period, our inversion of plasma temperature shows that the local plasma heating is gender. Thus we suggest that in the first half of our observation period, our inversion from H$\alpha$ line, namely 8 - 13 kK, is a reliable result.

The H$\alpha$ surge and filament formation can be linked to each other by the injection scenario proposed by \citet{chae03}. They both describe the process that cool and dense materials are erupted by the chromospheric reconnection. The differences are, most surges are erupted and materials flow along a magnetic arch without dips, thus they will drop back to chromosphere. But if the mass flows through a magnetic tube with dip, e. g., a filament channel, it will stay in this dip and become filament mass. Some of the works support this scenario \citep{liu05,zou16,zou17,wang18}. But most of the reported filaments are active region filaments or intermediate filaments, which have strong magnetic environment for surges to replenish enough materials to enough height. But for quiescent filaments, they have weaker magnetic environments, higher heights, whether this scenario still works is doubt.

In the simulation of \citet{jiang11}, they found the initial eruption of a surge is generated by the tension force of reconnection and the subsequent ascending motion is sustained by the pressure gradient force caused by plasma heating. Because of the weak magnetic field strength, the initial LOS velocity of our observation is about 11 - 13 km s$^{-1}$. Associating with the local heating, the LOS velocity decreases with an acceleration of 0.017 km s$^{-2}$ along the fibril (a resultant acceleration of gravity and pressure gradient force). Deriving from the initial doppler velocity and the acceleration, the height of this surge is about 4800 km, which is lower than the typical active region filament height. But such low height is due to two factors, one is the acceleration and the other is inclination angle of magnetic field lines. Evaluating from the initial velocities of this surge, the inclination angle of this surge is 62$^{\circ}$. Assuming that, the magnetic tubes of this surge have smaller inclination angle, such as 50$^{\circ}$, the plasma can be lifted up to about 10 Mm, which is the typical active region filament height. It means, even the magnetic reconnection occurring in weak magnetic field and causing a gender plasma heating, it still can replenish materials for active region filaments. But, even we assume that the magnetic tube is vertical, which means the initial velocity against the gravity is the resultant velocity of this surge, e. g., 31 km s$^{-1}$, the maximum height derived from this velocity and acceleration is 26 Mm. It means this surge can hardly replenish materials for those quiescent filaments with 25 Mm or even higher, since the initial velocity caused by weak magnetic field is too low. Especially, some footpoints of quiescent filaments are rooted in weaker magnetic field. Thus we think injection model does not work well in the formation of quiescent filaments.

\begin{acknowledgements}
The authors are grateful to the referee for the useful comments. This work was supported by the National Natural Science Foundation of China (41731067), Shenzhen Technology Project JCYJ20170307150645407, and the Specialized Research Fund for State Key Laboratories. W.C. acknowledges the support of the US NSF (AGS-0847126 and AGS-1250818) and NASA (NNX13AG14G).
\end{acknowledgements}

\label{lastpage}

\end{document}